\documentclass[12pt]{article}
\usepackage{graphicx}
\title{{\normalsize
~\\[-2.5cm]\hfill LU TP 99-19\\[-2mm]\hfill hep-ph/9907514}\\[0.5cm]
I : Chiral Perturbation for Kaons\\
II: The $\Delta I=1/2$-rule in the Chiral Limit}

\author{Johan Bijnens\\Department of Theoretical Physics 2, Lund University\\
S\"olvegatan 14A, S22362 Lund, Sweden}

\begin{document}
\maketitle
\begin{abstract}
I : Chiral Perturbation Theory is introduced and its applications
to semileptonic and nonleptonic kaon decays are discussed.

II: The method of large $N_c$ is used to
calculate $K\to\pi\pi$ nonleptonic matrix elements, in particular
the matching procedure between long and short-distance evolution
that takes all scheme dependence correctly into account is discussed.
Numerical results reproduce the  $\Delta I=1/2$
rule without the introduction of any free parameters.
\end{abstract}

\section{Introduction}

Chiral Perturbation Theory (CHPT)
is a very large subject now so I will only discuss
it briefly and then review the present status of its use in semileptonic
and nonleptonic kaon decays.
It has had several major successes in rare decays which are discussed in
the contribution by Isidori \cite{isidori}. The application
to $K^0_L\to\pi^+\pi^- e^+e^-$ is treated by Savage
\cite{savage}. As described in the
second part
and in several other talks \cite{hambye,other},
it is also very relevant in calculations of the
nonleptonic matrixelements.
In Section \ref{chiral} I very briefly describe the underlying
principles. The next section reviews the application to
kaon semileptonic decays, this is one of the main playgrounds for
CHPT and the area of some major successes. The use in
kaon decays to pions is then discussed in Sect. \ref{nonleptonic}.
We treat the use of CHPT in simplifying matrixelement
calculations in Sect. \ref{kpi}, predictions for $K\to3\pi$
in Sect. \ref{k3pi},
and chiral limit cancellations in $B_6$ in Sect. \ref{B6}.

Section \ref{dI=1/2} constitutes part II of this talk. Here I describe how
the large $N_c$ method can take into account the scheme dependence of
short-distance operators and first results\cite{BP}.
 
\section{Chiral Perturbation Theory}
\label{chiral}

CHPT grew out of current algebra where
systematically going
beyond lowest order was difficult. The use of effective Lagrangians
to reproduce current algebra results was well known and Weinberg
showed how to use them for higher orders \cite{weinberg}.
This method was improved and systematized
by Gasser and Leutwyler in the classic papers \cite{GL1} and
proving that CHPT is indeed the low-energy limit for QCD using only
general assumptions only was done by Leutwyler \cite{leutwyler}.
Recent lectures are \cite{CHPTlectures}.

The assumptions underlying CHPT are:
\begin{itemize}
\item
Global Chiral Symmetry and its spontaneous breaking to the vector subgroup:
$SU(3)_L\times SU(3)_R\to SU(3)_V$.
\item
The Goldstone Bosons from this spontaneous breakdown are the only relevant
degrees of freedom, I.e. the only possible singularities.
\item Analyticity, causality, cluster expansion and special relativity
\cite{leutwyler}.
\end{itemize}
The result is then a systematic expansion in meson masses, quark-masses,
momenta and external fields. The external field method allows to
find the minimal set of parameters consistent with chiral symmetry and the
rest is basically only unitarity. With current algebra and dispersive methods
it is in principle also possible to obtain the same results
but the method of Effective Field Theories is much simpler.

So for any application of CHPT two questions should be answered:
\begin{enumerate}
\item Does the expansion in momenta and quark masses converge ?
\item If higher orders are important then:\\
\begin{minipage}{12cm}
$\bullet$ Can we determine all the needed parameters from the data ?\\
$\bullet$ Can we estimate them if not directly obtainable.
\end{minipage}
\end{enumerate}

\section{Semileptonic decays}
\label{semileptonic}

The application of CHPT to semileptonic decays has been reviewed
in \cite{daphne} and in \cite{Orsay}. Since then first results
at order $p^6$ have appeared.
The situation order by order is:
\begin{itemize}
\item[$
{\cal L}_2$] 
  2 parameters: $F_0$, $B_0$ (+quark masses).
\item[${\cal L}_4$] : 10+2 parameters \cite{GL1}
7 are relevant; 3 more appear in the meson masses. In addition we also have the
Wess-Zumino term and one-loop contributions.
\item[${\cal L}_6$] : 90+4 parameters \cite{BCE2,FS}. In addition there
are two-loop diagrams and one-loop diagrams with ${\cal L}_4$ vertices.
\end{itemize}

\subsection{General Situation}

\mbox{\boldmath${ p^2}$} { Current Algebra} : sixties\\
\mbox{\boldmath$p^4$} {One-loop} 80's, early 90's\\
\mbox{\boldmath$p^6$} \begin{minipage}[t]{12cm}
$\bullet$  Estimates using {dispersive and/or models}: ``done''\\
$\bullet$    {Double Log Contributions}: 
 mostly done \cite{BCE1}.\\
$\bullet$   {Two-flavour full calculations}: Done.\\
$\bullet$   {Three flavour full} calculations: 
 few done, several in progress.
\end{minipage}\\
\mbox{\boldmath$e^2 p^2$} In progress.\\
{\bf Experiment} Progress from
DAPHNE, NA48, BNL, KTeV, \ldots

\subsection{$K_{l2}$}
\label{kl2}

These decays are used to determine $F_K$ and test lepton universality
by comparing $K\to\mu\nu$ and $K\to e\nu$. $F_\pi$ is similarly
determined from $\pi\to \mu\nu$.
The theory is now known to NNLO fully in CHPT \cite{ABT} (for $F_\pi$ also
\cite{KG})
The results are shown in Table \ref{tab:fpi} when the contributions
from the $p^6$ Lagrangian are set to zero, i.e. $C_i^r=0$,
 at the scale indicated.
The numbers in brackets are the extended double log approximation
of \cite{BCE1}. The inputs are
$10^3 L_{i=4,10}=(-0.3,1.4,-0.2,0.9,6.9,5.5)$;
$\mu$ = 0.77 GeV unless otherwise indicated
and for set A $10^3 L_{i=1,3}=(0.4,1.35,-3.5)$
while for set B $10^3 L_{i=4,10}=(-0.3,1.4,-0.2,0.9,6.9,5.5)$.
\begin{table} 
{ }\hfill
\begin{tabular}{c|ccc}
 & $F_\pi/F_0$ & $F_K/F_\pi$ & $(F_K/F_\pi)^{(6)}$\\
\hline
$p^2$ & 1 & 1 & --- \\
$p^4$ & 1.07 & 1.22 & ---\\
$p^6$ set A & 0.96 (1.08) & 1.27 (1.30) & 0.05 (0.08) \\
$p^6$ set A $\mu$=0.9 GeV & 0.96 (1.10) & 1.30 (1.34) & 0.08 (0.12) \\
$p^6$ set B & 0.90 (1.02) & 1.25 (1.28) & 0.035 (0.06) \\
\hline
\end{tabular}
\hfill{ }
\caption{\label{tab:fpi} Results for the ratios of $F_\pi$, $F_K$ and the
decay-constant in the chiral limit. The size of the $p^6$ only is in the
last column.}
\end{table}
We see that the variation with the $p^4$ input is sizable and that
the extended double log approximation gives a reasonable first estimate
for the correction.

\subsection{$K_{l2\gamma}$}

In this decay there are two  form-factors.
The axial form-factor is known to $p^4$\cite{Don,BEG}
and a similar calculation for $\pi\to e\nu\gamma$\cite{BT} shows a 25\%
correction and a small dependence on the lepton invariant mass $W^2$.
The vector form-factor is known to $p^6$\cite{Ametller} and has a 10 to 20
\% correction in the relevant phasespace.
The main interest in these decays is that it allows to test the anomaly and
its sign as well as the $V-A$ structure of the weak interactions.

\subsection{$K_{l2ll}$}

In these decays there are three vector and one axial form-factor.
The vector ones are known to $p^4$ \cite{BEG} and the axial
one to $p^6$ \cite{Ametller}. Especially the decays with $e\nu_e$
in the final state are strongly enhanced over Bremsstrahlung. Since
my last review \cite{Orsay} there is a new limit from BNL E787 \cite{bnle787}
of $B(K^+\to e^+\nu\mu^+\mu^-)\le 5§\cdot 10^{-7}$. All data are in good
agreement with CHPT.

\subsection{$K_{l3}$}

These decays, $K^{+,0}\to\pi^{0,-} \ell^+\nu$,
are our main source of knowledge of
the CKM element $V_{us}$. It is therefore important to have as
precise predictions as possible.
The form-factors
\begin{equation}
\langle\pi(p^\prime)|V_\mu^{4-i5}|K(p)\rangle = \frac{1}{\sqrt2}
\left[(p+p^\prime)_\mu f_+(t)+(p-p^\prime)_\mu f_-(t)\right]
\end{equation}
are usually parametrized by
$
f_+(t) \approx f_+(0)\left[1+\lambda_+{t}/{m_\pi^2}\right]
$
and 
$f_0(t) \equiv f_+(t)+{t f_-(t)}/({m_K^2-m_\pi^2})
 \approx f_+(0)\left[1+\lambda_0 {t}/{m_\pi^2}\right]$.

The CHPT calculation at order $p^4$ fits these parametrizations well\cite{GL3}.
The agreement with data is quite good except for the scalar slope
where there is disagreement between different experiments.
The extended double log calculation\cite{BCE1} has small quadratic slopes,
$\lambda^\prime_+$ and $\lambda^\prime_0$,
and small corrections to the linear slopes.
This, as shown in Table \ref{tab:kl3}, is good news
for improving the precision of $V_{us}$. $f_+$ is shown
for $K^0\to\pi^- e^+\nu$ where isospin breaking is smallest.
\begin{table}
{ }\hfill
\begin{tabular}{c|cccc}
& $p^2$ &  $p^4$ & Ext. double log & Experiment \\
\hline
$f_+$ & 1 & $-0.0023$ & $-0.005\to 0.004$ & input for $V_{us}$\\
$\lambda_+$ & 0 &  0.031 & $-0.006\to -0.0044$ & $0.029\pm 0.002$\\
$\lambda_0$ & 0 & 0.017 & $0.003\to 0.009$ & $0.025\pm 0.006$ $K^0_{\mu3}$\\
 &&&& $0.006\pm 0.007$ $K^+_{\mu3}$\\
$\lambda^\prime_+$ & 0 & small & $0.0002\to 0.0003$\\
$\lambda^\prime_0$ & 0 & small & $0.0001\to 0.0002$\\
\hline
\end{tabular}
\hfill{ }
\caption{\label{tab:kl3}CHPT and experimental results for $K_{l3}$ decays.}
\end{table}

\subsection{$K_{l3\gamma}$}

These decays have been calculated in CHPT to $p^4$ in \cite{BEG}.
There are 10 formfactors and after a complicated interplay between
all the various terms the final corrections to tree level are
small even though individual form factors have large corrections.
E.g. first adding tree level, then $p^4$ tree level and finally
$p^4$ loop level contributions changes $B(K^+_{e3\gamma})$
with $E_\gamma\ge 30$~MeV and $\theta_{\ell\gamma}\ge 20^o$
from $2.8~10^{-4}$ via $3.2~10^{-4}$ to $3.0~10^{-4}$.
Notice that $F_K/F_\pi = 1.22$ so agreement with tree level at the
10\% level is a good test of CHPT at order $p^4$.

Recent new results of $B(K^0_{e3\gamma}) =(3.61\pm0.14\pm0.21)~10^{-3}$
\cite{leber}(NA31) and $B(K^0_{\mu3\gamma})=(0.56\pm0.05\pm0.05)~10^{-3}$
\cite{bender} (NA48) are in good agreement with the theory results\cite{BEG}
of $(3.6\to4.0\to3.8)~10^{-3}$ and $(0.52\to0.59\to0.56)~10^{-3}$ respectively.
The three numbers correspond to the contributions included as above.

\subsection{$K_{l4}$}

In these decays, $K\to\pi\pi\ell\nu$, there are four form-factors,
$F,G,H,R$ as defined in \cite{daphne,BCG}. The $R$ form factor
can only be measured in $K_{\mu4}$ decays and is known to $p^4$\cite{BCG}.
$F$ and $G$ were calculated to $p^4$ in \cite{kl4} and improved using
dispersion relations in \cite{BCG}. The main data come from \cite{rosselet}
($K\to\pi^+\pi^- e^+\nu$)
and \cite{makoff} ($K_L\to\pi^\pm\pi^0e^\mp\nu$). The form factors were
parametrized as $X = X(0)(1+\lambda(s_{\pi\pi}/(4m_\pi^2)-1))$ with
the same slope for $X=F,G,H$. $H(0)= -2.7\pm0.7$ is a test of the anomaly
in both sign and magnitude, see \cite{Ametller} and references therein.
The other numbers are the main input for $L_1^r$, $L_2^r$ and $L_3^r$.
In table \ref{tab:kl4} I show the tree level results,
which expression, the $p^4$ or dispersive
improved, to determine $L_{1,2,3}^r$ of set A and B given in Sect.
\ref{kl2}, and the extended double log estimate of $p^6$\cite{BCE1}.
The results of the latter show similar patterns as the dispersive
improvement.
The full $p^6$ calculation is in progress and if the results are as indicated
by the extended double log approximation a refitting of $p^4$ constants
will be necessary. This is important since in these decays and
in pionium decays the $\pi\pi$ phaseshifts will be measured accurately
and their main theory uncertainty is the values of these constants.
A useful parametrization to determine these phases from $K_{l4}$
can be found in \cite{AB} as well as  further relevant references.
\begin{table}
{ }\hfill
\begin{tabular}{c|ccc}
 & $F(0)$ & $\lambda F(0)$ & $G(0)$ \\
\hline
$p^2$ & 3.74 & --- & 3.74       \\
$p^4$ & \multicolumn{3}{l}{set B; fit to $L_1,L_2,L_3$}\\
dispersive &  \multicolumn{3}{l}{set A; fit to $L_1,L_2,L_3$}\\
$p^6$ set A & 0.86& 0.38  & $-$0.15\\
$p^6$ set A $\mu=0.9~GeV$ & 1.13 & 0.51 & $-$0.20\\
$p^6$ set B & 0.75 & 0.17 & $-0.04$\\
experiment & $5.59\pm0.14$ & $0.45\pm0.11$ & $4.77\pm0.27$\\
\hline
\end{tabular}
\hfill{ }
\caption{
\label{tab:kl4} CHPT, dispersive, partial $p^6$ \cite{BCE1}
and experimental results for $K_{l4}$.}
\end{table}

\section{Nonleptonic decays}
\label{nonleptonic}

For rare decays see \cite{isidori}, here only
$K\to 0,\pi,\pi\pi,\pi\pi\pi$ are discussed.
The lowest order Lagrangian contains three
terms with parameters $G_8$, $G_{27}$, $G_8^\prime$ in
 the notation of \cite{BPP}.
The term with $G_8^\prime$, the  weak mass term, contributes to processes with
photons at lowest order and otherwise at NLO\cite{BPP}.
The NLO lagrangian contains about 30 parameters for the octet, denoted
by $E_i$, and twenty-seven, denoted by $D_i$, representation
of $SU(3)_L$ \cite{KMW1,EKW}.

\subsection{$K\to\pi,K\to 0 \leftrightarrow K\to\pi\pi$}
\label{kpi}

As shown in \cite{BPP} the method of \cite{wiseetal} can be
extended to $p^4$ using well defined off-shell Green functions of
pseudo-scalar currents. Except for one $E_i$ and one $D_i$
all the necessary ones can be obtained from $K\to\pi$
transitions\footnote{Using $K\to 0$ allows to obtain two more constants
than given in \cite{BPP}.}. To $K\to\pi\pi$ at order $p^4$ 7 $E_i$ and
6 $D_i$ contribute in addition to the three couplings of lowest order.
Of these 16 constants we can determine 14 from the much simpler
$K$ to $\pi$ and vacuum transitions. This allows thus a more stringent
test of various models than possible from on-shell $K\to\pi\pi$ alone.
Models like factorization etc. will probably be
needed in the foreseeable future to go to $K\to3\pi$ and various rare decays.

\subsection{CHPT for $K\to\pi\pi$ and $K\to\pi\pi\pi$}
\label{k3pi}

These decays were calculated to $p^4$ \cite{KMW2},
relations between them clarified in \cite{KDetal}
and some $p^6$ estimates to them were performed in \cite{AI}.

The main problem is to find experimental relations after all
parameters are counted. To order $p^2$ we have 2(1) and to $p^4$
7(3). The number in brackets refers to the $\Delta I=1/2$ observables
only. 
As observables (after using isospin)
we have 2(1) $K\to\pi\pi$ rates and 2(1)(+1) $K\to\pi\pi\pi$ rates.
We have 3(1)(+3) linear and 5(1)(+5) quadratic slopes.
The (+i) indicates the phases, in principle also measurable
and predicted but not counted here.
12 observables and 7 parameters leave five relations to
be tested. The fits and results are shown in Table \ref{tab:k3pi}
where we have also indicated which quantities are related.
See \cite{KDetal,MP} for definitions and references.
The new CPLEAR\cite{cplear} data improve the precision slightly.
$K\to\pi\pi$ rates are always input.
\begin{table}
{ }\hfill
\begin{tabular}{c|ccc}
\mbox{variable} & $p^2$ & $p^4$ &\mbox{experiment} \\
\hline
$\alpha_1 $ & $ 74 $ & $\mbox{input}(1) $ & $ 91.71\pm0.32$\\
$\beta_1 $ & $ -16.5  $ & $ \mbox{input}(2) $ & $ -25.68\pm0.27$\\
$\zeta_1 $ & $- $ & $ -0.47\pm0.18 $(1) & $ -0.47\pm0.15$\\
$\xi_1 $ & $- $ & $  -1.58\pm0.19 (2)$ & $ -1.51\pm0.30$\\
$\alpha_3 $ & $ -4.1 $ & $ \mbox{input}(3) $ & $ -7.36\pm0.47$\\
$\beta_3 $ & $ -1.0 $ & $ \mbox{input}(4) $ & $ -2.42\pm0.41$\\
$\gamma_3 $ & $ 1.8  $ & $ \mbox{input}(5) $ & $ 2.26\pm0.23$\\
$\xi_3 $ & $-$& $  0.092\pm0.030(4) $ & $ -0.12\pm0.17$\\
$\xi_3^\prime $ & $- $ & $ -0.033\pm0.077(5) $ & $ -0.21\pm0.51$\\
$\zeta_3 $ & $- $ & $ -0.011\pm0.006 $(3) & $ -0.21\pm0.08 $\\
\hline
\end{tabular}
\hfill{ }
\caption{\label{tab:k3pi}The predictions and experimental results
for the various $K\to3\pi$ quantities. Numbers in brackets refer
to the related quantities.}
\end{table}
It is important to tests these relations directly, the agreement at present
is satisfactory but errors are large.

CP-violation in $K\to3\pi$ will be very difficult to detect.
The strong phases needed to interfere with are very small, see \cite{AI}
and references therein. E.g. $\delta_2-\delta_1$ in $K_L\to\pi^+\pi^-\pi^0$
is predicted to be $-0.083$ and the experimental result is only
$-0.33\pm0.29$. Asymmetries are expected to be of order $10^{-6}$
so we can only expect to improve limits in the near future.

\subsection{$B_6$ in the chiral limit}
\label{B6}

In the usual definitions of $B_i$ factors in nonleptonic decays
\begin{equation}
B_6 \equiv \frac{\langle \mbox{out} | Q_6 | \mbox{in} \rangle}
{\langle\mbox{out}|Q_6|\mbox{in}\rangle_{\mbox{factorized}}}
\end{equation}
the denominator needs to be well defined.
This is {\em not} true for $B_6$ in the chiral limit.
The factorizable denominator contains the scalar radius which
is infinite in full chiral limit. This can be seen
in the CHPT calculation\cite{BP}.
\begin{equation}
G_8\bigg|_{Q_6\mbox{fact}} = -\frac{80 C_6(\mu) B_0^2(\mu)}{3F_0^2}
\left[L_5^r(\nu)-\frac{3}{256\pi^2}\{2\ln\frac{m_L}{\nu}+1\}\right]
\end{equation}
Here  $\nu$ is the CHPT scale and $m_L$ the meson mass, we can see that 
$G_{8fact}\longrightarrow \infty$ for $m_L\to 0$.

The nonfactorizable part has precisely the same divergence so that in the
sum it cancels.
Thus when calculating $B_6$ care must be taken to calculate factorizable
and nonfactorizable consistently so this cancellation that is required
by chiral symmetry takes place and does not inflate final results.

\section{The $X$-boson method and the $\Delta I=1/2$ rule in
the chiral limit.}
\label{dI=1/2}

In this section I shortly describe how in the context of the large $N_c$
method \cite{hambye,other,BBG} after the improvements of the momentum
routing\cite{BBG2} also the scheme dependence\cite{two-loops1,two-loops2}
can be described.
Other relevant references to the problem of nonleptonic matrix elements
are \cite{buras}.

The basic underlying idea is that we have more experience in hadronizing
currents. We therefore replace
the effect of the local operators of
$H_W(\mu) = \sum_i C_i(\mu) Q_i(\mu)$ at a scale $\mu$
by the exchange of a series of colourless $X$-bosons at a low scale $\mu$.
Let me illustrate the procedure in a simpler case of only one operator
and neglecting penguin contributions.
In the more general case all coefficients become matrices.
\begin{equation}
C_1(\mu)(\bar s_L\gamma_\mu d_L)(\bar u_L\gamma^\mu u_L)
\Longleftrightarrow
X_\mu\left[g_1 (\bar s_L\gamma^\mu d_L)+g_2 (\bar u_L\gamma^\mu u_L)
\right]\,.
\end{equation}
Colour indices  inside  brackets are summed over.
To determine $g_1$, $g_2$ as a function of $C_1$ we
set matrix elements of $C_1 Q_1$ equal to the equivalent ones
of $X$-boson exchange. This must be done  at a $\mu$
such that perturbative QCD methods can still be used and thus we can use
external states of quarks and gluons.
To lowest order this is simple. The tree level diagram
\begin{figure}
\includegraphics[width=\textwidth]{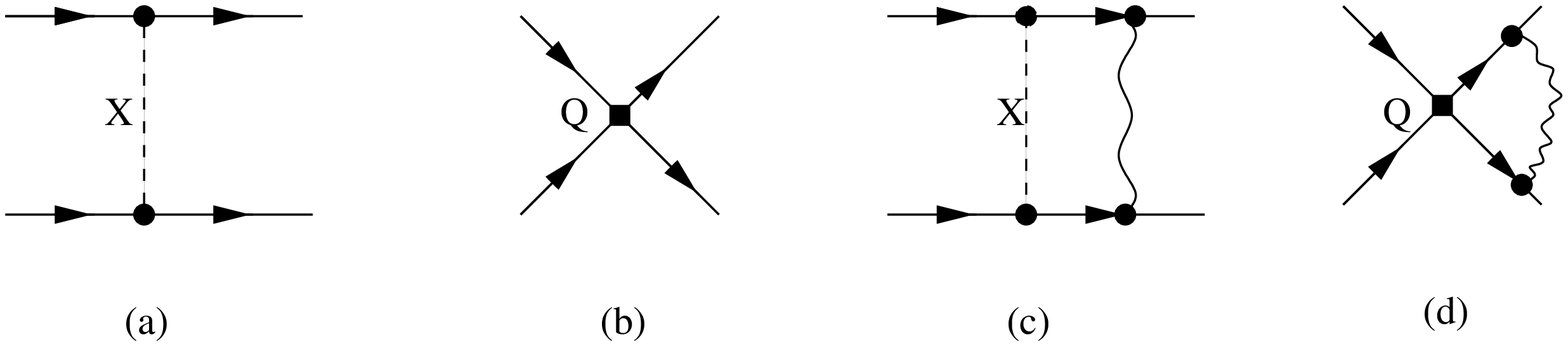}
\caption{\label{figX} The diagrams needed for the identification
of the local operator $Q$ with $X$-boson exchange in the case of
only one operator and no Penguin diagrams. The wiggly line
denotes gluons, the square the operator Q and the dashed line
the $X$-exchange. The external lines are quarks.}
\end{figure}
from Fig. \ref{figX}(a) is set equal to that of Fig. \ref{figX}(b)
leading to 
$
C_1 = {g_1 g_2}/{M_X^2}\,.
$
At NLO diagrams 
like Fig. \ref{figX}(c)
and \ref{figX}(d) contribute as well leading to
\begin{equation}
C_1\left(1+\alpha_S(\mu)r_1\right)
= \frac{g_1 g_2}{M_X^2}\left(1+\alpha_S(\mu)a_1+
\alpha_S(\mu)b_1\log\frac{M_X^2}{\mu^2}\right)\,.
\end{equation}
The left-hand-side (lhs)
is scheme-independent. The right-hand-side can be calculated in a
very different renormalization scheme from the lhs.
The infrared dependence of $r_1$ is
present in  precisely the same
way in $a_1$ such that $g_1$ and $g_2$ are scheme-independent
and independent of the precise infrared definition of the external state
in Fig. \ref{figX}.

One step remains, to calculate the matrix element
of $X$-boson exchange between meson external states.
The integral over $X$-boson momenta we split in two
\begin{equation}
\label{split}
\int_0^\infty dp_X\frac{1}{p_X^2-M_X^2}
\Longrightarrow
\int_0^{\mu_1}dp_X\frac{1}{p_X^2-M_X^2}
+\int_{\mu_1}^\infty dp_X\frac{1}{p_X^2-M_X^2}\,.
\end{equation}
The second term involves a high momentum that needs to flow back
through quarks or gluons and leads through diagrams like the one
of Fig. \ref{figX}(c)
to a four quark-operator with a coefficient
\begin{equation}
\frac{g_1 g_2}{M_X^2}\left(\alpha_S(\mu_1)a_2
+\alpha_S(\mu_1)b_1\log\frac{M_X^2}{\mu^2}\right)\,.
\end{equation}
The four-quark operator
needs to be evaluated only in leading order in $1/N_c$.
The first term in (\ref{split})
we have to evaluate in a low-energy model with as much QCD input as possible.
The $\mu_1$ dependence cancels between the two terms in (\ref{split})
if the low-energy model is good enough. 
The coefficients $r_1$, $a_1$ and $a_2$ give the
correction to the factorization used
in previous $1/N_c$ calculations.

It should be stressed that in the end all dependence on $M_X$ cancels
out. The $X$-boson is a purely technical device to correctly
identify the four-quark operators in terms of well-defined products of
nonlocal currents.

\subsection{Numerical results}

We now use the $X$-boson method with $r_1$ as given in \cite{two-loops1}
and $a_1=a_2=0$, the calculation of the latter is in progress,
and $\mu=\mu_1$. For $B_K$ we can extrapolate to the pole both
for the real case ($\hat B_K$) and in the chiral limit ($\hat B_K^\chi$).
For $K\to\pi\pi$ we can get at the
values of the octet ($G_8$), weak mass term ($G_8^\prime$)
and 27-plet ($G_{27}$) coupling.
We obtain $\hat B_K^\chi= 0.25\mbox{--}0.4\,;$
\begin{equation}
\hat B_K = 0.69\pm0.10\,;~
G_8= 4.3\mbox{--}7.5\,;~G_{27}=0.25\mbox{--}0.40\mbox{ and }
G_8^\prime=0.8\mbox{--}1.1\,.
\end{equation}
The experimental values are
$G_8\approx6.2$ and $G_{27}\approx0.48$ \cite{BP,KMW2}.

\begin{figure}
\begin{minipage}[t]{0.485\textwidth}
\includegraphics[width=\textwidth]{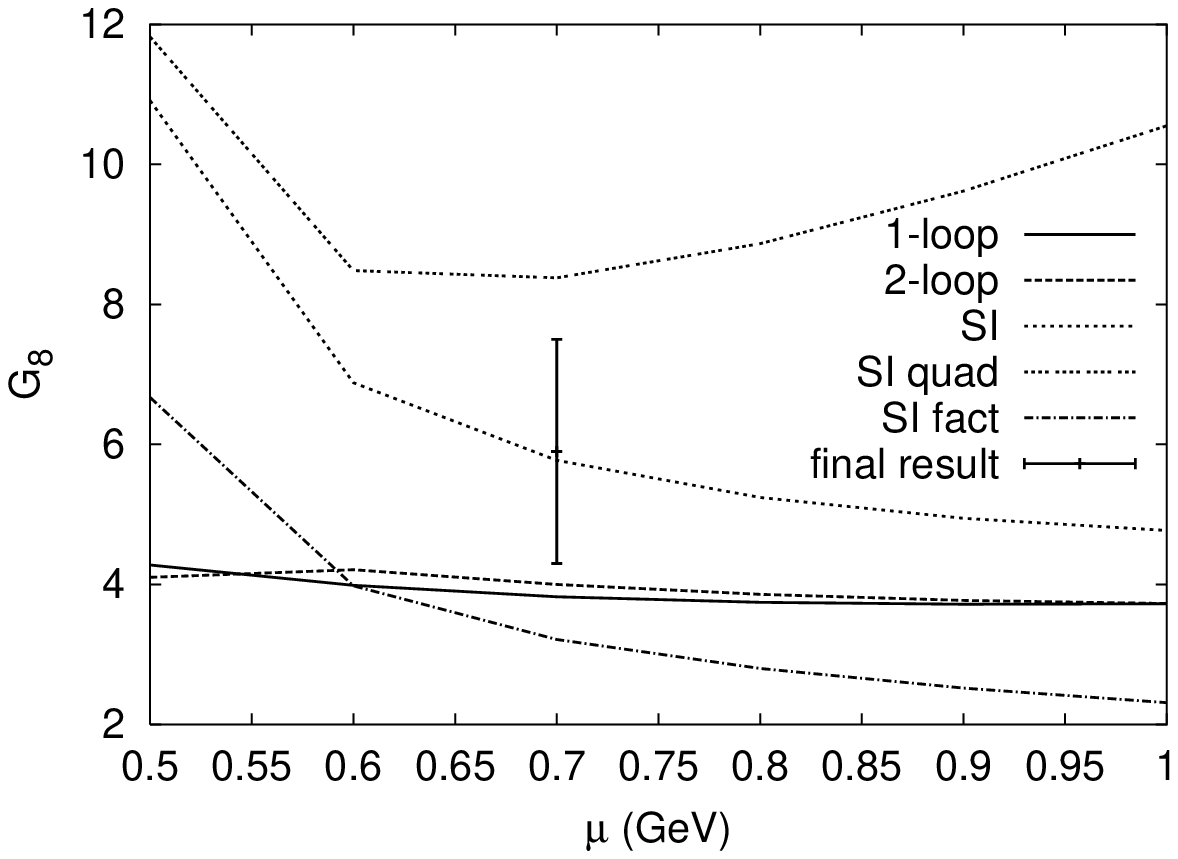}
\caption{\label{figg8} $G_8$ as a function of
$\mu$ using the ENJL model and Wilson coefficients at one-loop,
at 2-loop with and without the $r_1$ (SI). 
The factorization (SI fact)
and the approach of \cite{hambye} (SI~quad) are shown for SI also.}
\end{minipage}
\hfill
\begin{minipage}[t]{0.485\textwidth}
\includegraphics[width=\textwidth]{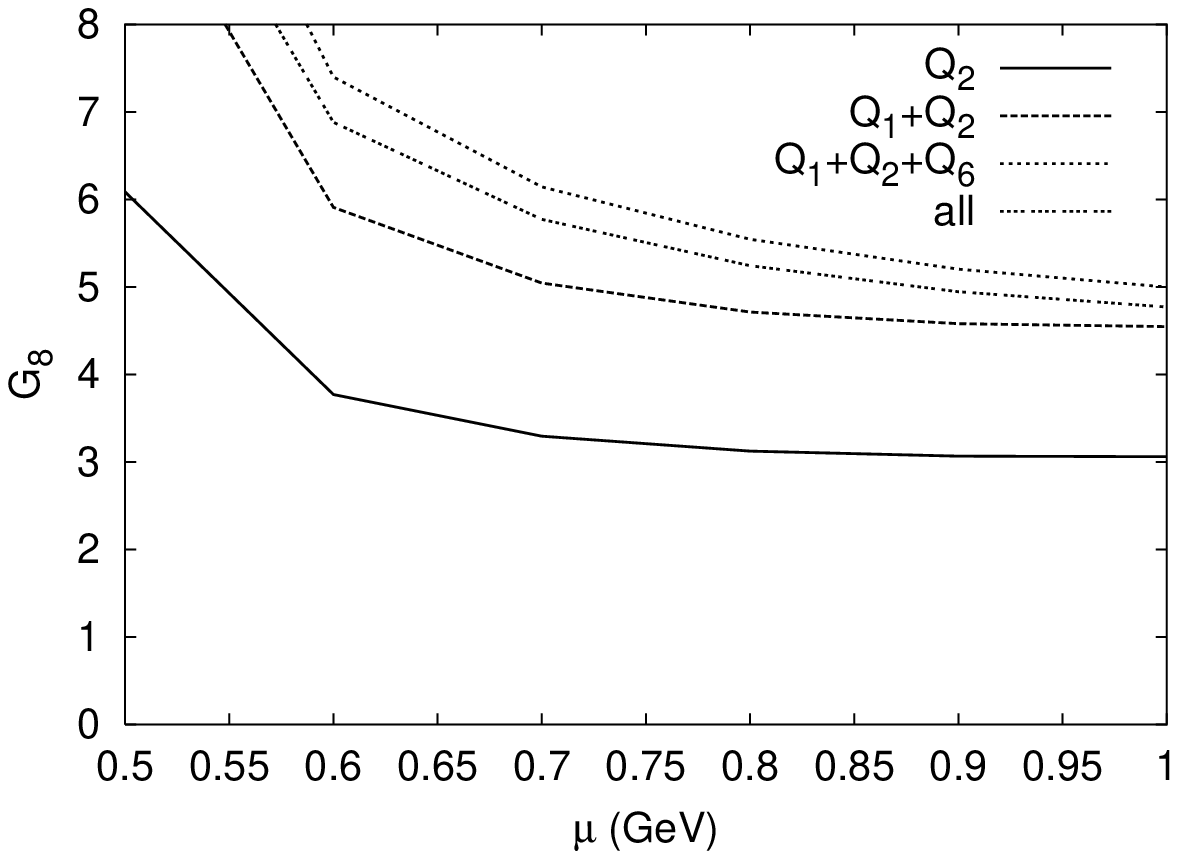}
\caption{\label{figg8_comp} The composition of $G_8$ as a function of
$\mu$. Shown are $Q_2$, $Q_1+Q_2$, $Q_1+Q_2+Q_6$ and all 6 $Q_i$.
The coefficients $r_1$ are included in the Wilson coefficients.}
\end{minipage}
\end{figure}

In Fig. \ref{figg8} the $\mu$ dependence of $G_8$ is shown
and in Fig. \ref{figg8_comp} the contribution from the various
different operators.
If we look inside the numbers we see that $B_6$ defined with only
the large $N_c$
term in the factorizable part, is about 2 to 2.2 for $\mu$ from
0.6 to 1.0~GeV.

\section{Conclusions}

CHPT is doing fine in kaon decays, especially in the semileptonic
sector where several calculations at $p^6$ are now in progress.
In the nonleptonic sector it provides several relations in $K\to3\pi$
decays. Testing these is an important part since it tells us how well
$p^4$ works in this sector. CHPT can also help in simplifying and identifying
potentially dangerous parts in the calculations of nonleptonic matrix elements.

The large $N_c$ method allows to include the scheme dependence appearing in
short-distance operators and when then all long-distance constraints
from CHPT and some other input are used encouraging results are obtained
for $K\to\pi\pi$ decays in the chiral limit.

\end{document}